\documentstyle[twocolumn,aps]{revtex}
\begin{document}
\input epsf
% \draft command makes pacs numbers print
\draft
% repeat the \author\address pair as needed
\title{\bf Is there a 4.5 PeV neutron line in the cosmic ray spectrum?}
\author{Robert Ehrlich}
\address{Physics Department, George Mason
University, Fairfax, VA 22030}
\date{\today}
\maketitle

\begin{abstract}

Recently we presented a model to fit the cosmic ray spectrum using the 
hypothesis that the electron neutrino is a tachyon.  The model predicted 
the existence of a neutron flux in the cosmic rays in a narrow region 
centered on $E = 4.5 \pm 2.2$ PeV. The published literature on
Cygnus X-3 reveals just such a $6\sigma$ spike of neutral
particles centered on E = 4.5 PeV.  A second prediction of the model
concerning integrated neutron fluxes at several energies also is 
consistent with published data.  A specific further test of the model is
proposed.

\end{abstract}
% insert suggested PACS numbers in braces on next line
\pacs{PACS: 14.60.St, 14.60.Pq, 95.85.Ry, 96.40.De}

% figures follow here
%
% Here is an example of the general form of a figure:
% Fill in the caption in the braces of the \caption{} command. Put the label
% that you will use with \ref{} command in the braces of the \label{} command.
%
% \begin{figure}
% \caption{}
% \label{}
% \end{figure}

% tables follow here
%
% Here is an example of the general form of a table:
% Fill in the caption in the braces of the \caption{} command. Put the label
% that you will use with \ref{} command in the braces of the \label{} command.
% Insert the column specifiers (l, r, c, d, etc.) in the empty braces of the
% \begin{tabular}{} command.
%
% \begin{table}
% \caption{}
% \label{}
% \begin{tabular}{}
% \end{tabular}
% \end{table}

%
% ****** End of file template.aps ******

%\begin{document}

\section{Introduction}

Recently we presented a model to fit the high energy cosmic ray spectrum 
using the hypothesis that the electron neutrino is a 
tachyon.\cite{Ehrlich}  A good fit to the spectrum was obtained using 
$|m_\nu| \equiv \sqrt{-m^2} = 0.5\pm 0.25$ eV/c$^2.$  The signature
prediction of the model is the existence of a neutron flux `spike' in 
the cosmic rays centered on $E = 4.5 \pm 2.2$ PeV, and having a width 
$\Delta\log E = 0.1$ (FWHM).  Although the existence of neutral cosmic
rays from point sources remains a highly controversial subject, we 
report here that an examination of the published literature on cosmic 
rays from Cygnus X-3 reveals just such a hitherto unreported neutral 
particle spike centered on E = 4.5 PeV with a level of statistical 
significance of $6\sigma.$  An additional prediction of the model that 
the integrated flux of neutrons above 0.5 EeV should be 0.048 percent 
that above 2 PeV is also consistent with results from two out of three 
experiments.

Although few physicists have taken tachyons seriously since they were 
first proposed in 1962\cite{Bilaniuk}, their existence is clearly an
experimental question. In 1985 Chodos, Hauser and 
Kosteleck\'{y}\cite{Chodos85}, suggested that neutrinos were tachyons --
an idea that is consistent with experiments used to determine the
neutrino mass. Chodos et al.\cite{Chodos92,Chodos94} also suggested a 
remarkable empirical test of the tachyonic neutrino hypothesis, namely 
that stable particles should decay when they travel
with sufficiently high energies.  Consider, for example, the 
energetically forbidden decay $p\rightarrow n + e^++ \nu_e.$  In order to
conserve energy in the CM frame the neutrino would need to have
$E < 0$.  But tachyons with $m^2 <0$ have $E<p,$ and therefore the
sign of their energy in the lab frame $E_{lab} = \gamma(E + \beta p 
cos\theta)$ will be positive for a proton velocity $\beta > \beta_{th} 
\equiv -E/p cos\theta.$  With the aid of a little kinematics it can 
easily be shown that the threshold energy for proton decay is 
$E_{th}\approx 1.7|m_{\nu}|^{-1}$ PeV, with $|m_{\nu}|$ in eV.

Thus, if neutrinos are tachyons, energetically forbidden decays become 
allowed when the parent particle has sufficient energy -- in seeming 
contradiction with the principle of relativity that whether or not a 
process occurs should not depend on the observer's reference frame.  
That contradiction is only an apparent one, however, because what 
appears to the lab observer as a proton decay emitting a neutrino 
appears to the CM observer as a proton absorbing an antineutrino from a 
background sea.

\section{Cosmic Rays}

Since cosmic rays bombard the Earth with energies far in excess of what 
can be achieved in present day accelerators, it is natural to ask 
whether any evidence for a process such as proton decay exists there at 
very high energies. One striking feature of the cosmic ray spectrum is 
the ``knee" or change in power law that occurs at $E \approx 4$ PeV. 
Various two-source mechanisms have been suggested to account for this spectral
feature, but some researchers have identified it as arising from a single
type of source.\cite{Erlykin}  In 1992 Kosteleck\'{y}\cite{Kostelecky} 
suggested that for a tachyonic neutrino mass $|m_\nu| \approx 0.3 eV,$ 
the proton decay threshold energy occurs at the knee of the cosmic ray 
spectrum, and could explain its existence.  The idea is that cosmic ray 
nucleons on their way to Earth would lose energy through a chain of 
decays $p\rightarrow n\rightarrow p\rightarrow n\rightarrow \cdots,$ 
which would deplete the spectrum at energies above $E_{th}.$  However, 
Kosteleck\'{y} regarded the existence of the knee by itself as 
insufficient evidence for the tachyonic neutrino hypothesis in view of 
other more conventional explanations of the knee of the cosmic ray 
spectrum.  He also did not attempt to model the spectrum, nor mention 
the signature neutron spike.

Recently this author has developed a tachyonic neutrino model that fits 
a number of features of the cosmic ray spectrum in addition to the 
knee.\cite{Ehrlich}  These include the existence and position of the 
``ankle" (another change in power law at $E \approx 6$ EeV),  the
specific changes in power law at the knee and ankle, the changes in 
composition of cosmic rays with energy, and the ability of cosmic rays 
to reach us above the conjectured GZK ``cutoff."\cite{Greisen,Takeda} 
Although the fit to the cosmic ray spectrum was a good one, the model is 
highly speculative, because it is at variance with
conventional wisdom about cosmic rays and it arbitrarily assumed
that the decay rate for protons (for $E>E_{th}$) was far greater than 
that for neutrons.

Nevertheless, the model did make the striking prediction of a cosmic ray 
neutron flux in a narrow range of energies just above $E_{th}$ -- a 
neutron ``spike."  The pile up of neutrons in a narrow interval just 
above $E_{th}$ is a consequence of the fractional energy loss of the 
nucleon in proton decay becoming progressively smaller, the closer the 
proton energy gets to $E_{th}$. The position of the predicted cosmic ray 
neutron spike depends on the value assumed for $|m_\nu|$.  From the fit 
to the cosmic ray spectrum we found $|m_\nu| = 0.5 \pm 0.25$ eV/c$^2,$ and
hence we predicted a neutron spike at $E = 4.5 \pm 2.2$ PeV.  In fact
the model predicted that most nucleons should be neutrons for
$E > E_{th},$ because it was assumed that as nucleons lose energy in the 
$p\rightarrow n\rightarrow p\rightarrow \cdots$ decay chain, the
lifetime and hence the decay mean free path for neutrons is far
greater than for protons, and so nucleons above $E_{th}$ would spend 
nearly all of their time en route as neutrons.\cite{assumption}  But, the model also
predicts that for energies above the spike the neutron component does 
not become an appreciable fraction of the total cosmic ray flux until 
around 1 EeV.  While neutrons might reach Earth at EeV energies in 
conventional cosmic ray models, it would be difficult to understand any 
sizable neutron component at energies as low as E=4.5 PeV, where the
neutron mean free path before decay would be only about 100 ly.  In the 
present model, however, A = 1 cosmic rays can travel very many neutron 
decay lengths and still arrive as neutrons because many steps of the 
$p\rightarrow n\rightarrow p\rightarrow \cdots$ decay chain occur for
nucleons having energies above $E_{th}.$

\section{Cygnus X-3 Data}

One way to look for a neutron flux would be to find a cosmic ray signal 
that points back to a specific source, since neutrons are unaffected by 
galactic magnetic fields.  Starting in 1983 a number of cosmic ray 
groups did, in fact, report seeing signals in the PeV range from
Hercules X-1 and Cygnus X-3.  At the time these signals were
believed to be either gamma rays or some hitherto unknown long-lived 
neutral particle, since neutrons, as already noted, should not live long 
enough to reach Earth (except in the present model).  Some
of the experiments coupled detection of extensive air showers with 
detection of underground muons.\cite{Samorski,Marshak}  The observed 
high muon intensity was found to be consistent with hadrons but not with 
showers induced by gamma rays.\cite{Marshak,Stanev} It was widely 
believed that the mass of the neutral particle was $m\approx 1 
$ GeV/c$^2.$\cite{Cudell} Thus, all the observed or conjectured 
properties of these particles were consistent with neutrons: neutral 
strongly interacting particles with $m\approx 1$ GeV/c$^2.$

Following a period of excitement in the 1980's, many researchers began 
to look critically at some of the observations of ultra-high energy 
cosmic rays from point sources.  This skepticism was based in part on 
the inconsistencies between results reported in different experiments.  
As Chardin and Gerbier have noted\cite{Chardin}, a number of papers used
data selection procedures that made direct comparisons difficult, e.g., 
using different phase intervals to make cuts, variously reporting the 
total flux or only the flux in a particular phase bin, and reporting 
only ``muon-poor" events.  Also, some papers appeared to
inflate the statistical significance of their results.

But, the most serious challenge to the idea of neutral particles in the 
PeV range from Cygnus X-3 and other point sources came from a trio of 
high sensitivity experiments\cite{Alexandreas,Aglietta,Cronin} that 
reported seeing no signals from point sources claimed earlier. In the 
most sensitive experiment of the three, the upper limit on the flux of 
neutral particles from Cygnus X-3 above 1.175 PeV was far below the 
fluxes reported by those experiments claiming signals 
earlier.\cite{Borione}  There seems to be only two possibilities: either 
{\it all} the earlier experiments claiming signals were in error, or 
Cygnus X-3 and other reported sources all had turned off about the time 
improved instrumentation became available.  Table I offers some support 
for the latter possibility, because (a) the phases of the signals are in 
rough agreement in three experiments, and (b) the integrated flux above 
a PeV does appear to systematically decrease over time taking all 
experiments together.  (Among those claiming signals only those claiming 
more than $4\sigma$ have been listed, and among those citing upper 
limits only those giving upper limits on the flux above a PeV have been 
listed.)  The suggestion that signals from Cygnus X-3 have fallen with 
time was first raised by N. C. Rana et al. based on X-ray and gamma ray 
data in four different wavelength regions.\cite{Rana}  In what follows, 
we make the ``optimistic" assumption that earlier experiments were 
seeing real signals, and we consider to what extent those reports of 
signals from Cygnus X-3 support the prediction of a 4.5 PeV neutron 
spike.

%REWRITE THE FOLLOWING AFTER I HEAR FROM ONG:
%
%However, our prediction that the flux from point sources consists of a
%neutron spike at 4.5 PeV could explain how Cygnus X-3 and other sources
%can be just as luminous as ever, and yet not have registered in the more
%sensitive experiments.  Let us assume that the signal from
%Cygnus X-3 is a 4.5 PeV spike, and that it occurs at a particular
%orbital phase as had been reported in most experiments claiming a
%signal.  Also, let us assume that the integrated backround flux from
%other background sources has a $E^{-1.7}$ power law characteristic of
%cosmic rays below the knee.  If one merely looks at a phase plot for all
%events pointing back to Cygnus X-3 for which $E > E_0,$ then the
%background to signal ratio scales as ${E_0}^{-1.7},$ and one's ability
%to see a spike in the phase plot seriously degrades for values of $E_0$
%much below 4.5 PeV.  For example, the backrgound to signal ratio is
%$10^{1.7} = 50$ times worse for $E_0 = 0.1 PeV$ than $E_0 = 1 PeV.$  One
%can get much better detection sensitivity by searching for phase peaks
%in seperate energy bins.  Although, Borione et al. did in fact look at
%phase plots sorted according to energy bins, they apparently only looked
%up to 1.175 PeV, above which the statistics were considered insufficient
%-- which could account for their missing a signal consisting of a spike
%at 4.5 PeV. But, the preceding explanation of the failure of a high
%sensitivity experiment to see a signal from point sources hardly
%constitutes evidence for a neutron flux spike -- how can we see the
%effect directly?

In the 1980's there were eight cosmic ray groups that cited fluxes in 
the PeV range of signals pointing back to Cygnus X-3, (some which were 
inconsistent as mentioned earlier.)  In nearly all cases limited 
statistics required reporting the flux integrated over energy in only 
one or at most two energy intervals.  

\begin{table}[hbt]
\begin{center}
\begin{tabular}{ r r r r r r}
Ref & Years & E in PeV & Flux & Stat. sig. & Phase\\
\hline
\cite{Samorski} &  76-79 & $>2$ & $7.4\pm 3.2$ & $4.4\sigma$ & 0.1-0.3 \\
\cite{Hayashida}&  78-81 & $>1$ & $<3$         &             &         \\
\cite{Lloyd}    &  79-82 & $>3$ & $1.5\pm0.3$  & $5\sigma$   & 0.225-0.25\\
\cite{Muraki}   &  86-88 & $>1$ & $2.7\pm0.5$  & $4.7\sigma$ & 0.25-0.30\\
\cite{Cronin}   &  89    & $>1$ & $<23$        &             &        \\
\cite{Borione}  &  90-95 & $>1.175$ & $<0.1$   &             &        \\
\end{tabular}
\end{center}
\caption{Experiments reporting integrated fluxes (or upper 
limits) in units of $\times 10^{-14}$ particles cm$^{-2}$ sec$^{-1}$ for
Cygnus X-3 for PeV energies. Only experiments reporting nonsporadic 
signals claimed to be at a level of more than $4\sigma$ have been 
listed.  The van der Klis and Bonnet-Bidaud ephemeris has been used for 
finding the phase interval in each case.} \end{table}

One group (Lloyd Evans et al.\cite{Lloyd}), however, had good enough 
statistics to report fluxes in eight energy bins spanning the location 
of the predicted 4.5 PeV neutron spike, and it had an energy acceptance 
threshold near $E_0$ = 1 PeV, which could give one energy bin before the 
spike itself. The signal seen by Lloyd Evans et al. from Cygnus X-3 did 
not appear until the data is selected on the basis of orbital phase 
determined from the X-ray binary's 4.79 h orbital period, and the time 
of signal arrival.  Lloyd-Evans et al. found that if they looked at the 
number of counts in 40 phase bins, one of these bins showed a sizable 
excess (73 counts when the average was 39). The information in Table II
is taken from Lloyd-Evans et al.\cite{Lloyd}, with the last column added by
this author.  Fig. 1 displays the data in that last column.  We would 
expect a flat distribution on the basis of chance, assuming that the 
signal were just a statistical fluctuation.  In fact, averaged over all 
phases, the distribution must be flat and zero height, regardless of 
whether the signal is real or not.  Note, that a spike appears centered 
on the value predicted by the tachyonic neutrino model, and that all the 
remaining bins have a flux consistent with zero.  The gaussian curve 
drawn with arbitrary height in the figure shows what would be predicted 
by the model given a neutron spike of width $\Delta\log E = 0.1 (FWHM)$ 
and a 50 percent energy resolution ($\Delta\log E = \pm 0.176$).
According to Lloyd-Evans, the actual resolution
was probably around 50 percent, and very likely less than 100
percent\cite{informal}  We estimate the statistical
significance of this spike occurring by chance by dividing the excess 
number of events in the two bins straddling 5 PeV by the square root of 
the expected number of events in those two bins: 
$28.4/\sqrt{22.6}=6.0\sigma.$  It is interesting that in their article, 
Lloyd-Evans et al. displayed only the integrated flux $I(>E)$ versus 
energy, and hence failed to mention the spike.  Instead, they simply 
noted that the integrated spectrum appeared to steepen right after 10 
PeV.

How can we be sure that the spike seen in Lloyd-Evans et al. data is not an
artifact of the data analysis or a statistical fluctuation? Six standard 
deviations
may seem interesting, but the original peak in their phase plot was far
less impressive, particularly allowing for a ``trials factor" of 40, 
since such a peak might have been seen in any one of the 40 phase bins.
Suppose that in fact the original peak in the phase plot were a 
statistical fluctuation, how could one then get a $6\sigma$ peak in the 
flux versus energy distribution for events in a specific phase bin?  
Clearly, such a peak would require some correlation between energy and 
phase.  This could in principle occur, because observed cosmic ray 
energy is correlated with declination angle, and hence with time of day.  
However, all cosmic rays in a given phase bin arrive at one of five, 
i.e., 24/4.79, times throughout the day, and those arrival times slowly 
advance from day to day, since the Cygnus X-3 period is not exactly 
divisible into 24 hours.  Thus, over the years of data-taking each phase 
bin would sample times of the day with an almost uniform distribution, 
making it difficult to see how a phase-energy correlation could occur.

\begin{table}[hbt]
\begin{center}
\begin{tabular}{ r r r r}
E (in PeV) & Observed & Expected & Excess $\pm\sqrt{Expected}$\\
\hline
1-3             &  16      & 13.9     & 2.1$\pm$ 3.7 \\
3-5             &  34      & 16.4     & 17.6$\pm$ 4.0 \\
5-11            &  17      & 6.2      & 10.8$\pm$ 2.5 \\
11-18           &  4       & 2.4      & 1.6$\pm$ 1.6 \\
18-36           &  3       & 4.3      & $-$1.3$\pm$ 2.1 \\
36-72           &  6       & 3.4      & 2.6$\pm$ 1.8 \\
72-140          &  2       & 0.8      & 1.2$\pm$ 0.9 \\
$>$140          &  0       & 0.6      & $-$0.6$\pm$ 0.8 \\
\end{tabular}
\end{center}
\caption{Observed and expected event counts reported by Lloyd-Evans et
al. in differential energy bins for the phase interval 0.225-0.250.  The 
last column has been added by the author. The ``Expected" counts for 
each energy interval are based
on the average over all phases.}
\end{table}

\begin{figure}[hbt]
\begin{center}
\leavevmode
\epsfxsize=3.25in
\epsffile{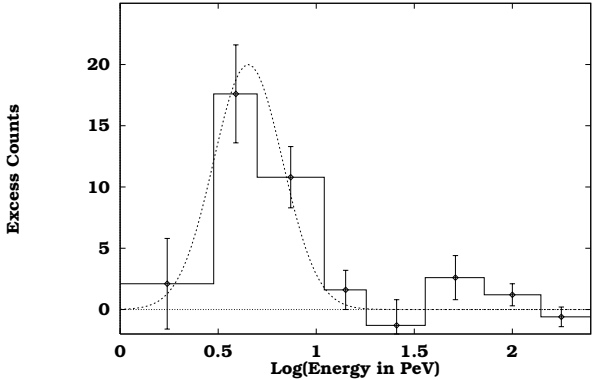}
\caption{Data points from the last column of Table II plotted at the
middle of each interval in log E in PeV.  The gaussian curve
centered on 4.5 PeV is what one would expect to find in Lloyd-Evans 
data, given a neutron spike of width $\Delta\log E = 0.1$(FWHM), and a
50 percent energy resolution ($\Delta\log E = \pm 0.176$)}

\end{center}
\end{figure}

(It could be that at their source 
the phase and energy of cosmic rays are correlated, but in that case we would
be dealing with a real source, not a statistical fluctuation, as 
hypothesized above.)

Ideally, one would want to combine the Lloyd-Evans et al. data with that 
of other experiments in the PeV region to see if
the spike either is destroyed or enhanced.  Several problems arise with 
the other existing data, in which a signal is claimed from Cygnus X-3: 
one experiment used only ``muon-poor" events\cite{Kifune}, two
experiments reported only the integral flux above some energy (no energy 
bin defined)\cite{Kirov,Baltrusaitis}, two reported the flux in an
energy bin three times the width used by 
Lloyd-Evans\cite{Samorski,Tonwar}, and none was contemporaneous with
Lloyd-Evans, thereby severely diminishing their utility.

%Of course, if there
%really is a 4.5 PeV neutron spike, one should be able to see it in the
%three high statistics experiments with a low energy acceptance
%thresholds as a peak in the phase distribution, merely by first
%excluding events with $E < 1 PeV$ before making the phase plot.

%What about the usefulness of existing data on point sources at sub-PeV
%energies? In the tachyonic neutrino model neutrons decay back and stay
%protons below 4.5 PeV, so that any observation of cosmic ray neutrons
%below this energy would be a clear contradiction.  However, it is quite
%possible that ultrahigh energy gamma rays would be produced along with
%cosmic rays, so the point sources below $E_{th}$ that have been seen
%might well be only gamma rays, or statistical fluctuations.

Aside from the spike, one other prediction of the tachyonic neutrino 
model is that neutrons should also be seen as a significant and rising 
fraction of the cosmic ray flux above around 1.0 EeV.   In fact, two 
cosmic ray groups have reported seeing neutral particles from Cygnus X-3 
having energies above 0.5 EeV with fluxes of $1.8\pm 0.7$\cite{Teshima}, 
and $2.0\pm 0.6$\cite{Cassiday}, while a third group reporting merely an 
upper limit to the flux $<0.4$\cite{Lawrence} -- all in units of 
$10^{-17}$ particles cm$^{-2}$ s$^{-1}$.

These measured fluxes above 0.5 EeV can be compared directly with the
neutron flux predictions from the tachyonic neutrino 
model.\cite{Ehrlich} As noted previously, the ratio of the integral flux 
of neutrons above 0.5 EeV to that above 2 PeV is predicted to be $R = 
4.8\times 10^{-4}.$  The predicted neutron flux for $E > 0.5$ EeV is
then $R$ times the measured flux reported by Lloyd-Evans et al. for $E > 
2$ PeV, or: $R\times 7.4\pm 3.2 \times 10^{-14} = 3.5 \pm
1.5 \times 10^{-17}$ particles cm$^{-2}$ s$^{-1}$, which is in quite 
good agreement with the two groups that measured a flux, rather than an 
upper limit.  Although subsequent data accummulation by these two groups 
failed to show a signal from Cygnus X-3\cite{informal1}, that only adds
additional support to the hypothesis that the source faded over time.

%Only a relatively small number of candidate cosmic ray point sources have so far been
%examined, but the total number of such sources may be very large.  For
%example, if Cygnus X-3 can be considered a source that provides a larger
%fraction of the total neutron flux received at Earth than the average
%source (most of which would presumably be at greater
%distances), we can compute an estimated lower limit on the number of
%sources, N, whose cosmic rays reach Earth from the ratio of the absolute
%flux our model predicts above 2 PeV to that reported by Lloyd-Evans for
%Cygnus X-3 to find: $N > 6 \times 10^4.$

If it is true that Cygnus X-3 and other point sources were active in the 
early 1980's and subsequently have turned off, is there any way to check
whether there really is a 4.5 PeV neutron spike without waiting for 
specific sources to come back on?  Without knowing where the sources
are, the model can make no prediction of the
anisotropy or the the angular
distribution of sources of high energy cosmic rays.  However, recall that the model
predicts that {\it all} the cosmic rays include a 4.5 PeV neutron spike, 
not just those pointing back to the handful of possible sources looked 
at so far.  Thus, if one selects events in a narrow energy band centered 
on 4.5 PeV, one could look at their arrival directions on the two 
dimensional map of the sky, and see if there is a noticeable clustering 
of points, which would indicate neutral particles coming from specific 
sources.  Moreover, if those sources were episodic, one should observe a 
nonuniform distribution in arrival times for events for a given source.

Consider a specific example.  The integrated flux in the 4.5 PeV spike 
is 0.1 neutrons per m$^2$-sr-s, which would give around 3 million counts 
over 5 years for an array of area 250,000 m$^2$.  If the array had an 
energy resolution of 100 percent, it would also record a background 
count rate roughly four times as great in the energy bin centered on 4.5 
PeV.  Suppose the angular resolution were $\Delta\theta = 0.01$ rad, which
would allow up to $4/{\Delta\theta}^2 = 4\times 10^4$ solid angle bins 
to be defined.  Each bin would then have on the average 400 background 
counts.  Further suppose that the cosmic rays reaching Earth came from N 
point sources, then those solid angle bins pointing back to sources 
would have an average signal to background ratio: $10^4/N.$ 
Identification of sources should then be possible, unless N were larger 
than the number of solid angle bins, and no subset of sources were
appreciably brighter than others.

\section{Summary}

In summary, a highly speculative tachyonic neutrino model\cite{Ehrlich}, 
which fits the cosmic ray spectrum well, predicts a spike of neutrons at 
an energy where, given the neutron lifetime and distance to likely 
sources, very few should appear.  A search through the literature for 
sources of neutral cosmic rays has identified a particular experiment 
with a favorable energy acceptance threshold, good enough statistics, 
and enough energy bins spanning the region of the neutron spike to test 
the prediction.  The data do show a $6\sigma$ spike located right at the 
predicted energy, which was not identified in the original work.  The 
failure of other subsequent more sensitive experiments to see a signal 
from Cygnus X-3 would seem to require that this source has since turned 
off -- a possibility given some support by both time trends of data from
different experiments, and data within the same experiments.  The 
characteristics of the neutral particles from Cygnus X-3 seem to be 
consistent with neutrons rather than gamma rays, based on muon data from 
various experiments.  For the EeV region, where the model also predicts 
neutrons (though not a spike), two out of three experiments show a 
positive signal from Cygnus X-3, and they report a flux whose magnitude 
(relative to the flux in the spike) is well-predicted by the model. The 
hypothesis that the electron neutrino is a tachyon would seem to be 
supported, and it can be further tested without waiting for specific 
point sources to come back on.

%One obvious check on the claim made in this paper would be to conduct
%another high statistics experiment looking at the region centered on the
%predicted spike for Cygnus X-3 and other likely point sources.  But, one
%could also do an experiment without requiring that cosmic rays point back to
%any identified source, rather one could use the cosmic rays themselves to find
%the point sources.  As long as the number of sources were not too great,
%one should be able to identify neutrons in terms of multiple cosmic rays
%pointing back to some set of fixed points.  One indication that this
%alternative method may not be feasible comes from the following very
%rough estimate of the number of point sources of cosmic rays.

\acknowledgements

The author wishes to thank John
Wallin for helpful comments.  He also could have not done without
the critical comments of his colleague Robert Ellsworth.


\begin{references}

\bibitem{Ehrlich} R. Ehrlich, Phys. Rev. D, {\bf 60}, (1999) 17302.

\bibitem{Bilaniuk} O. M. P. Bilaniuk, V. K. Deshpande, and E. C. G. Sudarshan,
Am.J.Phys. {\bf30} (1962) 718.

\bibitem{Chodos85} A. Chodos, A.I. Hauser, and V. A. Kosteleck\'{y}, Phys.
Lett. {\bf 150B} (1985) 295.

\bibitem{Chodos92} A. Chodos, V. A. Kosteleck\'{y}, R. Potting, and E. Gates,
Mod. Phys. Lett. A {\bf 7} (1992) 467.

\bibitem{Chodos94} A. Chodos, and V. A. Kosteleck\'{y}, Phys.Lett.B {\bf
336} (1994) 295.

\bibitem{Erlykin} A.D. Erlykin, and A.W. Wolfendale, Astropart. Phys. 
{\bf 8}, (1998) 265.

\bibitem{Kostelecky} V. A. Kosteleck\'{y}, in F. Mansouri and J.J. Scanio,
eds., Topics on Quantum Gravity and Beyond, World Scientific, Singapore, 
1993.

\bibitem{Greisen} K. Greisen, Phys. Rev. Lett. {\bf 16} (1966) 748;
G. T. Zatsepin and V. A. Kuz'min, JETP Lett. {\bf 4} (1966) 78.

\bibitem{Takeda} M. Takeda, et al., Phys. Rev. Lett. {\bf 81} (1998) 
1163.

\bibitem{assumption} Were it not for the assumption that the neutron 
mean decay length greatly exceeds that of protons, the magnetic 
deflection of protons would eliminate directional correlations, and make
the detection of neutrons based on their directionality highly problemmatic
for all but nearby sources.

\bibitem{Samorski} M. Samorski and W. Stamm, Astrophys. J. {\bf 268}
(1983) L17

\bibitem{Marshak} M.L. Marshak et al. Phys. Rev. Lett. {\bf 54} (1985) 
2079.

\bibitem{Stanev} T.S. Stanev, T.K. Gaisser and F. Halzen, Phys. Rev D 
{\bf 32} (1985) 1244.

\bibitem{Cudell} J.R. Cudell, F. Halzen and P. Hoyer, Phys. Rev. D, {\bf 
36} (1987) 1657.

\bibitem{Chardin} G. Chardin and G. Gerbier, Astron. Astrophys., {\bf
210} (1989) 52.


\bibitem{Alexandreas} Alexandreas et al., Astrophys. J. Lett. {\bf 383} 
(1991) L53.

\bibitem{Aglietta} M. Agietta et al., Astropart. Phys., {\bf 3} (1995) 
1.

\bibitem{Cronin} J.W. Cronin et al., Phys. Rev. D, {\bf 45} (1992) 4385.

\bibitem{Borione} A. Borione et al., Phys. Rev. D, {\bf 55} (1997) 1714.

\bibitem{Rana} N.C. Rana, M. Sadzinska, J. Wdowczyk, and A.W. Wolfendale,
Astron. Astrophys., {\bf 141}, (1984) 394.

\bibitem{Ciampa} D. Ciampa et al., Phys. Rev. D, {\bf 42} (1990) 281.

\bibitem{Lloyd} J. Lloyd-Evans et al., Nature, {\bf 305} (1983) 784.

\bibitem{Hayashida} N. Hayashida et al., Proc. Int. Conf. Cosmic Rays 
(Paris) (1981) XG 1.1-4.

\bibitem{Muraki} Y. Muraki et al., Astrophys. J. {\bf 373} (1991) 657.

%\bibitem{Li-Ma} T. Li and Y. Ma, Astrophys. J. {\bf 272} (1983) 317.

\bibitem{informal} Informal communication with Jeremy Lloyd-Evans.

\bibitem{Kifune}T. Kifune et al., Astrophys. J. {\bf 301} (1986) 230.

\bibitem{Kirov} I.N. Kirov, J. Phys. G, {\bf 18} (1992) 2027.

\bibitem{Baltrusaitis} R. M. Baltrusaitis et al., Astrophys. J. {\bf 
297} (1985) 145

\bibitem{Tonwar} S.C. Tonwar et al., Astrophys. J. Lett. {\bf 330} 
(1988) L107

%\bibitem{Groups} C.L. Bhat Astrophys. J., {\bf 306}
%(1986) 587; 


\bibitem{Teshima} M. Teshima et al., Phys. Rev. Lett., {\bf 64} (1990)
1628.

\bibitem{Cassiday} G.L. Cassiday et al., Phys. Rev. Lett., {\bf 62}
(1989) 383.

\bibitem{Lawrence} M. A. Lawrence, D. C. Prosser, and A.A Watson, Phys. 
Rev. Lett. {\bf 63} (1989) 1121.

\bibitem{informal1} Informal communications with M. Nagano of the Akeno
collaboration, and with Paul Sommers of the Fly's Eye collaboration.

\end{references}
\end{document}